\documentclass[aps,pra,showpacs,twocolumn,superscriptaddress,amsmath,amssymb]{revtex4}

\usepackage{amsmath,amsfonts,amssymb}
\usepackage{graphicx}

\usepackage[draft]{hyperref}
\usepackage{amsmath}
\newcommand {\startbild}{\begin{figure}}
\newcommand {\stopbild}{\end{figure}}
\newcommand {\staf}{\begin{equation}}
\newcommand {\stof}{\end{equation}}
\newcommand {\staffeld}{\begin{eqnarray}}
\newcommand {\stoffeld}{\end{eqnarray}}
\newcommand {\staa}{\begin{align}}
\newcommand {\stoa}{\end{align}}

\newcommand{\ket}[1]{|#1\rangle}

\renewcommand{\vec}[1]{{\bf #1}}

\begin{document}

\title{Measurement induced focussing of radiation from independent single photon sources}

\author{R.~Wiegner}
\affiliation{Institut f\"{u}r Optik, Information und Photonik, Universit\"{a}t Erlangen-N\"{u}rnberg, Erlangen, Germany}

\author{S.~Oppel}
\affiliation{Institut f\"{u}r Optik, Information und Photonik, Universit\"{a}t Erlangen-N\"{u}rnberg, Erlangen, Germany}
\affiliation{Erlangen Graduate School in Advanced Optical Technologies (SAOT), Universit\"at Erlangen-N\"urnberg, Erlangen, Germany}

\author{J. von Zanthier}
\affiliation{Institut f\"{u}r Optik, Information und Photonik, Universit\"{a}t Erlangen-N\"{u}rnberg, Erlangen, Germany}
\affiliation{Erlangen Graduate School in Advanced Optical Technologies (SAOT), Universit\"at Erlangen-N\"urnberg, Erlangen, Germany}

\author{G.~S.~Agarwal}
\affiliation{Department of Physics, Oklahoma State University, Stillwater, OK, USA}
\affiliation{Erlangen Graduate School in Advanced Optical Technologies (SAOT), Universit\"at Erlangen-N\"urnberg, Erlangen, Germany}

\date{\today}

\begin{abstract}
We present a technique based on multi-photon detection which leads to a strong focussing of photons scattered by independent single photon emitters. For $N$ single photon sources it is shown that if $m - 1$ photons are detected in a particular direction (with $m \leq N$) the probability to detect the $m$-th photon in the same direction can be as high as $100\,\%$. This measurement induced focussing effect is already clearly visible for $m>2$. 
\end{abstract}


\maketitle

The manipulation of the spatial radiation characteristics of light sources is an outstanding problem in quantum optics. The efficient directional emission of photons into well-defined modes by single photon sources is for example of vital importance for tasks in quantum information processing. In recent years significant progress has been made to obtain a higher focussing of radiation, either using geometrical approaches, e.g., collecting photons with optical devices like lenses \cite{Davidson04,Pereira08} or mirrors  \cite{Davidson08,Wineland09}, or exploiting  effects from cavity QED, i.e.,  
using microcavities \cite{Yamamoto02,Bouwmeester07}, photonic crystal waveguides \cite{Sandoghdar04}, photonic nano-wires \cite{Loncar10} or nano-antennas \cite{vanHulst10,Götzinger11}. 

An alternative approach to achieve a higher directionality in the emitted radiation is the use of entangled light sources \cite{Dicke54,Lukin03,Kimble03,Scully06,Cirac08}. Here, directionality is accomplished intrinsically without employing additional optical devices. 
A simple explanation of this phenomenon based on quantum interferences valid also for potentially widely separated sources has been proposed recently \cite{Wiegner11a}. 
However, the entanglement of a large number of emitters is still challenging, even though significant progress has been made recently \cite{Blatt05,Wineland05,Blatt11}. Therefore it would be desirable if an ensemble of independent emitters in a separable state 
would show the same spatial focussing effects as the entangled system.

In this paper we propose a measurement scheme which leads to a strong focussing of the incoherent radiation emitted by an ensemble of non-interacting uncorrelated single photon sources, e.g., atoms which are initially prepared in the fully excited state. The technique is based on multi-photon detection generating source correlations which produce the heralded peaked emission pattern \cite{comment0}.

\begin{figure}[h!]
\centering
\includegraphics[width=0.3 \textwidth]{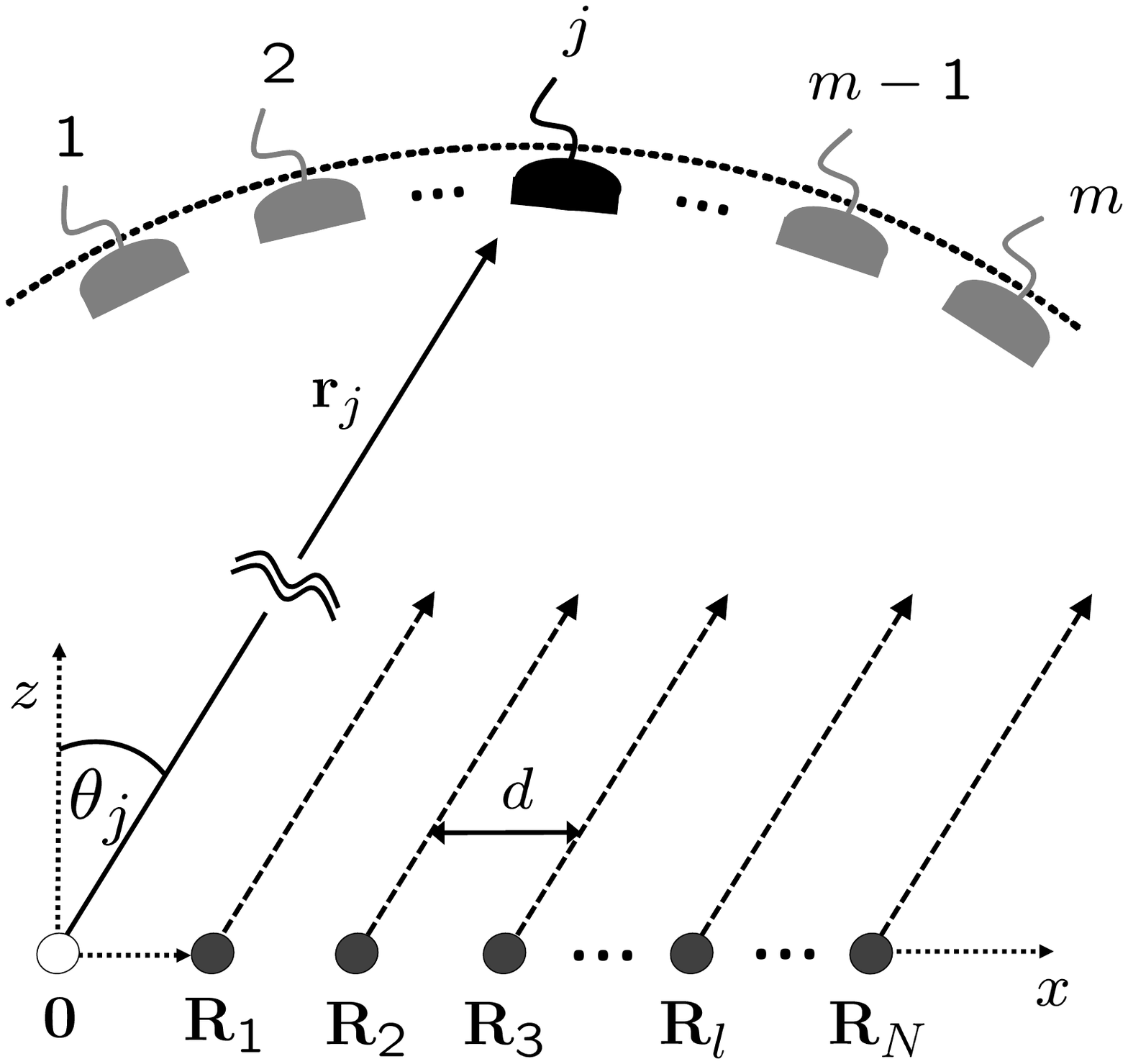}
\caption{\label{setup} Simplified scheme of the considered setup: $N$ two-level atoms located at positions ${\vec R}_{l}$ ($l \in {1, ..., N}$) along the $x$-axis and separated by a distance $d$ are initially fully excited. $m$ detectors are situated in the $x-z$ plane at ${\vec r}_j$ ($j \in {1, ..., m}$), each measuring a single photon in the far field of the atoms.}
\end{figure}

In the following we consider a chain of $N$ identical single photon sources, e.g., two-level atoms with upper level $\ket{e_{l}}$ and ground state $\ket{g_{l}}$, located at positions ${\vec R}_{l}$ ($l \in \{ 1,...,N\}$) along the $x$-axis (cf.~Fig.~\ref{setup}). We assume an equal spacing $d$ between the emitters and $kd > 1$ in order to neglect all atomic interactions like the dipole-dipole interaction, where $k=\frac{2\pi}{\lambda}$ is the wave number of the transition $\ket{e_{l}} \rightarrow \ket{g_{l}}$. 
We suppose that the atomic chain is initially fully excited to the separable state $\ket{S_{N}}$ with
\staf
\label{S-state}
\ket{S_{N}} \equiv \prod_{l = 1}^{N} \ket{e_l} \,.
\stof
We further assume $m$ detectors placed in the $x-z$ plane in the far field in a circle around the sources at positions ${\vec r}_j$ ($j \in \{ 1,...,m\}$, $m\leq N$), each supposed to detect a single photon. This process can be described by the $m$-th order correlation function 
\staffeld
\label{Eq1n}
G^{(m)}_{\ket{S_{N}}} (\vec{r}_1, t_1; ... ; \vec{r}_m, t_m)  \equiv \langle \hat{E}^{(-)}(\vec{r}_1,t_1) ...\, \hat{E}^{(-)}(\vec{r}_m,t_m) \nonumber\\ \otimes \, \hat{E}^{(+)}(\vec{r}_m,t_m) ...\,\hat{E}^{(+)}(\vec{r}_1,t_1) \rangle\,,
\stoffeld
where the positive frequency part of the electric field operator $\hat{E}^{(+)}(\vec{r}_j,t_j)$ is given by \cite{Agarwal74} 
\staf
\label{Eq2a}
\hat{E}^{(+)}(\vec{r}_j,t_j) \sim - \frac{e^{i\,k\,r_j}}{r_j^3} \sum_{l=1}^{N} \vec{r}_j \times\,(\vec{r}_j \times \vec{p}_{ge}) \, e^{-i\,\varphi_{lj}} \;\hat{s}^{-}_l(t_j) \,.
\stof
Here, $\hat{E}^{(-)} = \hat{E}^{{(+)}^\dagger}$, $\vec{p}_{ge}$ is the dipole moment of the transitions $\ket{e_{l}}\rightarrow\ket{g_{l}}$, $\hat{s}^{-}_l = |g_l\rangle\langle e_l|$ is the atomic lowering operator, and $\varphi_{lj}$ the optical phase difference accumulated by a photon emitted at $\vec{R}_l$ and detected at ${\vec r}_j$ relative to a photon emitted at the origin (cf.~Fig.~\ref{setup})
\staf
\label{Eq3}
\varphi_{lj} \equiv \varphi_{lj}({\vec r}_j,\vec{R}_l)  = - k \, \frac{\vec{r}_j\cdot\vec{R}_l}{r_j} = - l\,kd\,\sin\theta_j.
\stof
Note that according to Fig.~\ref{setup} $r_j$ is approximately equal for all $j$. 
In the following we consider for simplicity $\vec{p}_{ge}$ to be oriented along the $y$ direction, so that $\vec{p}_{ge}\cdot {\vec r}_j = 0$ . With these assumptions and omitting all constant factors, Eq.~(\ref{Eq2a}) takes the form
\staf
\label{Eq2}
\hat{E}^{(+)}(\theta_j,t_j) \sim \sum_{l=1}^{N} e^{-i\,\varphi_{lj}} \;\hat{s}^{-}_l (t_j)\,.
\stof
Here, the field is defined dimensionless and hence all correlation functions of $m$-th order would be dimensionless. The actual values can be obtained by multiplying $G^{(m)}_{\ket{S_{N}}}$ with $m$ times the intensity produced by a single atom.

Let us start by investigating the $m$-th order correlation function for the separable state $\ket{S_{N}}$. At equal times it calculates to
\staffeld
\label{GMGENSa}
\mathrm{G}^{(m)}_{\ket{S_{N}}}(\theta_1,...,\theta_m) = ||\sum_{\substack{\sigma_1, ..., \sigma_m = 1 \\ \sigma_1 \neq ... \neq \sigma_m}}^{N} \prod_{j = 1}^{m} e^{-i\,\varphi_{\sigma_j j}} \ket{g_{\sigma_j}}||^2 \nonumber \\
= \sum_{\substack{\sigma_1, ..., \sigma_m = 1 \\ \sigma_1 < ... <  \sigma_m}}^{N} |\sum_{\substack{\sigma_1,...,\sigma_m \\\in \,{\cal S}_m}}\prod_{j = 1}^{m} e^{-i\,\varphi_{\sigma_j j}}|^2\, ,
\stoffeld
where the norm of the state vector $\ket{\psi}$ is given by $||\ket{\psi}||^2 = \langle \psi | \psi \rangle$, the single bars $|...|$ abbreviate absolute values, and the expression $\sum_{\substack{\sigma_1,...,\sigma_m \\\in \,{\cal S}_m}}$ denotes the sum over the symmetric group ${\cal S}_m$ with $m$ elements $\sigma_1,...,\sigma_m$ and cardinality $m!$. 
The products $\prod_{j = 1}^{m} e^{-i\,\varphi_{\sigma_j j}}$ denote $m$-photon quantum paths where $m$ photons are emitted from $m$ sources at $\vec{R}_{\sigma_j}$ and recorded by the $m$ detectors at ${\vec r}_j$. Since, due to the far-field scheme none of the $m$ detectors can distinguish which of the $N \geq m$ atoms emitted a particular photon, we have to sum over all possible combinations of $m$-photon quantum paths, 
what is expressed by the sum $\sum_{\substack{\sigma_1, ..., \sigma_m = 1}}^{N}$ in the first line of Eq. (\ref{GMGENSa}). Hereby, the condition $\sigma_1 \neq ... \neq \sigma_m$ is applied since terms with $\sigma_i = \sigma_j$ vanish as $\hat{s}^{-}_{\sigma_j}\,\ket{g_{\sigma_j}} = 0$. Considering that several combinations of $m$-photon quantum paths lead to the same final atomic state and thus have to be superposed coherently, we end up with the modulus square in the second line of Eq.~(\ref{GMGENSa}). 
Hereby, for the $\binom{N}{m}$ different final atomic states, the corresponding transition probabilities $|...|^2$ have to be added incoherently, what results in the first sum $\sum_{\substack{\sigma_1, ..., \sigma_m = 1 \\ \sigma_1 < ... <  \sigma_m}}^{N}$ of the second line of Eq.~(\ref{GMGENSa}). 

Consider now that $m-1$ out of the $m$ detectors are placed at ${\vec r}_1$ and the last detector at ${\vec r}_2$. Under these conditions the $m$-th order correlation function takes the form
\staffeld
\label{GMGENSb}
\mathrm{G}^{(m)}_{\ket{S_{N}}}(\theta_1,...,\theta_1,\theta_2) = \frac{N!(m-1)!}{(N-m)!}\hspace{2cm}\nonumber\\
\times\left( \frac{N-m}{N-1} + \frac{m-1}{N(N-1)}\frac{\sin^2(N\frac{\varphi_{11}-\varphi_{12}}{2})}{\sin^2(\frac{\varphi_{11}-\varphi_{12}}{2})}\right),
\stoffeld
with a visibility ${\cal V} = (\mathrm{G}^{(m),\mathrm{max}}_{\ket{S_{N}}} - \mathrm{G}^{(m),\mathrm{min}}_{\ket{S_{N}}}) /$ $({\mathrm{G}^{(m),\mathrm{max}}_{\ket{S_{N}}} + \mathrm{G}^{(m),\mathrm{min}}_{\ket{S_{N}}}})$ given by
\staf
\label{vis1}
{\cal V} = \frac{m-1}{m+1 - \frac{2m}{N}} \, .
\stof
For $m = 1$ the visibility is zero, i.e., the mean radiated intensity is a constant, which illustrates the fact that the atoms scatter incoherently. For $1 < m \ll N$ the visibility is approximately given by ${\cal V} = \frac{m-1}{m+1}$, whereas for $m = N$ the maximum value ${\cal V} = 1$ is obtained .  

From Eq.~(\ref{GMGENSb}) it can be seen that even though all atoms emit incoherently, the $m$-th order correlation function $\mathrm{G}^{(m)}_{\ket{S_{N}}}(\theta_1,...,\theta_1,\theta_2)$ displays for $m>1$ the interference pattern of a coherently illuminated grating, with the central maximum located at $\vec{r}_2=\vec{r}_1$ having a width $\delta\theta_2$ (FWHM) of
\staf
\label{peakwidth}
\delta\theta_2 \approx \frac{2\pi}{N\,k\,d}\,.
\stof
For increasing numbers of emitters $N$ we thus observe a strong focussing of the radiated intensity in the direction of ${\vec r}_1$.
Yet, according to Eq.~(\ref{vis1}), the visibility of the $m$-th order correlation function decreases for growing $N$ and fixed $m$. Observing a strong focussing of the emitted radiation together with a visibility of 100~\% requires thus to measure $\mathrm{G}^{(m)}_{\ket{S_{N}}}(\theta_1,...,\theta_1,\theta_2)$ for large numbers of atoms $N$ and $m = N$. However, if a visibility ${\cal V} \approx \frac{1}{3}$ is considered sufficient, it suffices to measure $\mathrm{G}^{(2)}_{\ket{S_{N}}}(\theta_1,\theta_2)$ for $N \gg 2$ to observe a focussing close to a $\delta$-function, as ${\cal V} \xrightarrow{N \rightarrow \infty} \frac{1}{3}$. 

\begin{figure}[h!]
\centering
\includegraphics[width=0.45 \textwidth]{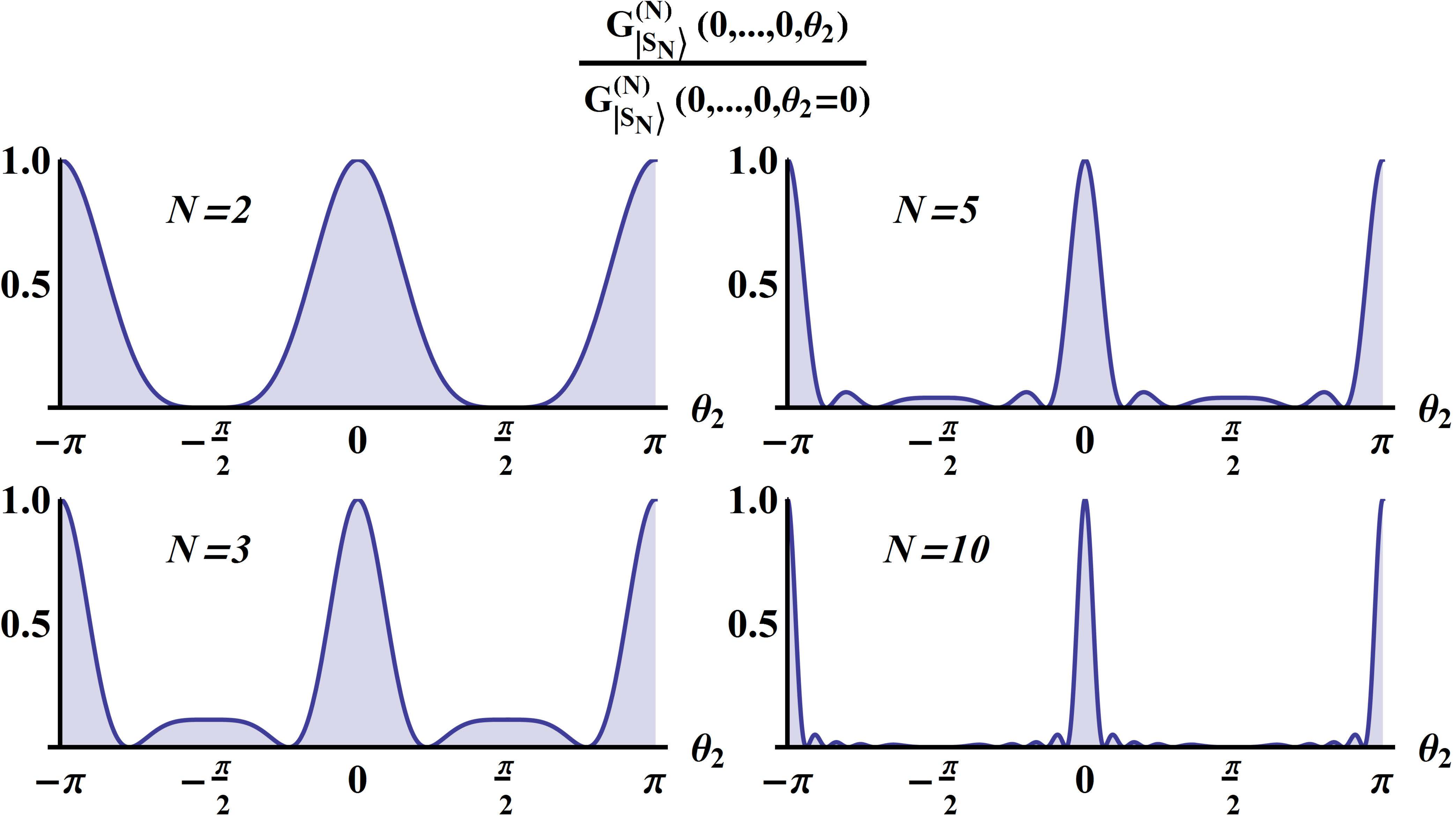}
\caption{\label{differentGN} Plots of the correlation function $\mathrm{G}^{(N)}_{\ket{S_{N}}}(0,...,0, \theta_2)$ for $N = \{2,3,5,10\}$ and $kd = \pi$, each normalized to its maximum value.}
\end{figure}

Figure~\ref{differentGN} displays $\mathrm{G}^{(N)}_{\ket{S_{N}}}(0,...,0, \theta_2)$ as a function of the observation angle $\theta_2$ for $N = \{2,3,5,10\}$. For a better comparison each function is normalized to its maximum value. Further, we chose $\theta_1 = 0$ and $kd = \pi$ to keep the focus to the central maximum. Clearly, the width of the $m$-th order correlation function is decreasing for increasing correlation order $m$ and number of atoms $N$.

We finally give a physical explanation for the discussed behavior, being quite unexpected for an ensemble of independently radiating single photon sources. Since the effect is strongest for $m=N$, let us focus on this case; a generalization to $N>m$ is straightforward. Consider the initially fully excited state $\ket{S_{N}}$ (Eq.~(\ref{S-state})). After detection of the first photon, this state is projected to 
\staffeld
\label{W-statein}
\ket{S_{N}}_{1} = \frac{1}{\sqrt{N}} \left( \ket{g_1}\ket{e_2}...\ket{e_N} + \ket{e_1}\ket{g_2}...\ket{e_N} \right.\nonumber \\  
\left. + ... + \ket{e_1}\ket{e_2}...\ket{g_N}\right) \,, \hspace{1cm}
\stoffeld
since due to the far-field assumption the photon could be scattered by any of the $N$ atoms. As more and more photons are recorded, the state of the atomic system evolves, so that, after $N-1$ detection events, the state $\ket{S_{N}}_{N-1}$ is attained given by 
\staffeld
\label{W-state}
\ket{S_{N}}_{N-1} = \frac{1}{\sqrt{N}} \left(\ket{e_1}\ket{g_2}...\ket{g_N} + \ket{g_1}\ket{e_2}\ket{g_3}...\ket{g_N} \right.\nonumber \\  
\left.+ ... + \ket{g_1}...\ket{g_{N-1}}\ket{e_N}\right) \,.  \hspace{1cm}
\stoffeld
This state corresponds to the single-excitation W-state $\ket{W_{1,N-1}}$ \cite{comment1}. In other words, employing a measurement scheme where $N-1$ photons are detected at position ${\vec r}_1$, we project the initially fully excited state $\ket{S_{N}}$ onto $\ket{W_{1,N-1}}$ \cite{comment2}. 
The spatial emission pattern of this state is then measured by the last detector at ${\vec r}_2$, corresponding to a measurement of the mean intensity $\mathrm{G}^{(1)}_{\ket{W_{1,N-1}}}(\theta_2)$ for the state $\ket{W_{1,N-1}}$. 
For this state a strong  focussing of the mean radiated intensity in the forward direction occurs what can be explained in a purely quantum path framework \cite{Wiegner11a}.
We thus showed that a property intrinsic to measurements on quantum systems can be used to produce a directional emission of radiation.

SO, JvZ and GSA thank the Erlangen Graduate School in Advanced Optical Technologies (SAOT) by the German Research Foundation (DFG) in the framework of the German excellence initiative for funding. RW~and~SO~gratefully acknowledge financial support by the Elite Network of Bavaria and the hospitality at the Oklahoma State University. This work was supported by the DFG.


\begin{thebibliography}{99}
\bibitem{comment1} Note that the individual terms in Eqs. (\ref{W-statein}) and (\ref{W-state}) could pick up different phase factors depending on the positions of the detectors and the locations of the atoms; see, e.g., \cite{Thiel07}.
\bibitem{comment2} We note that there are recent experiments on generating single-photon W-states \cite{Kimble09} or even storing atomic W-states due to projective measurement and dynamically allocating them in quantum interfaces \cite{Kimble10}.
\end{thebibliography}
\end{document}